\documentclass[12pt]{article}
\usepackage{amssymb,amsmath,epsfig}

\begin{document}

\title{\bf Nonlinear Electrodynamics in $f(T)$ Gravity and Generalized Second Law of Thermodynamics}
\author{M. Sharif \thanks {msharif.math@pu.edu.pk} and
Shamaila Rani\thanks {shamailatoor.math@yahoo.com}\\
Department of Mathematics, University of the Punjab,\\
Quaid-e-Azam Campus, Lahore-54590, Pakistan.}

\date{}

\maketitle
\begin{abstract}
In this paper, we study the nonlinear electrodynamics in the
framework of $f(T)$ gravity for FRW universe along with dust matter,
magnetic and torsion contributions. We evaluate the equation of
state and deceleration parameters to explore the accelerated
expansion of the universe. The validity of generalized second law of
thermodynamics for Hubble and event horizons is also investigated in
this scenario. For this purpose, we assume polelike and power-law
forms of scale factor and construct $f(T)$ models. The graphical
behavior of the cosmological parameters versus smaller values of
redshift $z$ represent the accelerated expansion of the universe. It
turns out that the generalized second law of thermodynamics holds
for all values of $z$ with event horizon for power-law scale factor
whereas it holds in a specific range of $z$ with Hubble horizon for
power-law and both horizons in polelike scale factors.
\end{abstract}
\textbf{Keywords:} $f(T)$ gravity; Magnetic field; Dark energy;
Generalized second law of thermodynamics.\\
\textbf{PACS:} 04.50.kd; 13.40.-f; 95.36.+x; 05.70.-a.

\section{Introduction}

The fact that the universe is expanding at every point in space has
become the most popular issue in cosmology. It is found that the
universe is nearly spatially flat and consists of about $74\%$ dark
energy (DE) (Perlmutter et al. 1997, 1998; Riess et al. 1998) and
the remaining $26\%$ corresponds to matter. Dark energy has positive
energy density with large negative pressure in order to derive the
acceleration of the universe. There are many proposals which serve
as a candidate of the DE in spite of lack of best fit model for this
acceleration. Modified theories of gravity (Nojiri and Odintsov
2007; Paul et al. 2009) has played an important role during last
decades to explain this accelerated expansion. The generalized
teleparallel theory of gravity (Bengochea and Ferraro 2009; Linder
2010; Yang 2011; Ferraro and Fiorini 2011; Myrzakulov 2011; Tsyba et
al. 2011) dubbed as $f(T)$ gravity is commonly used to explore the
insights of the universe with $T$ as the torsion scalar.

There are several cosmological ingredients in the universe including
radiations, dark matter and DE. The properties of these ingredients
are well specified by the equation of state (EoS) parameter $\omega$
which is the ratio of pressure to energy density of the universe.
The radiation dominated phase corresponds to $\omega=1/3$, whereas
$\omega=0$ represents the matter dominated phase. The DE dominated
phase inherits different regions with the help of EoS parameter
including the quintessence region for $-1<\omega <-1/3$, vacuum
energy due to the cosmological constant for $\omega=-1$ and phantom
region for $\omega<-1$. The EoS parameter for $f(T)$ gravity also
corresponds to these regions in different scenarios. Recently, we
have reconstructed the $f(T)$ models using EoS parameter for the
above mentioned cases and explored the accelerated expansion of the
universe (Sharif and Rani 2011a). Also, the relationship between
$f(T)$ gravity and k-essence model has been discussed with the help
of this parameter to present the evolving universe (Sharif and Rani
2011b).

Bamba et al. (2011) examined the EoS parameter in this gravity by
taking into account exponential, logarithmic and their combined
models which result different DE regions. Karami and Abdolmaleki
(2012) investigated the validity of generalized second law of
thermodynamics (GSLT) for Hubble horizon in $f(T)$ gravity using
power-law and exponential models. They concluded that GSLT holds for
both these models from early to present universe, while it is
violated in the future epoch. Bamba and Geng (2011) explored the
thermodynamics in equilibrium and non-equilibrium descriptions for
apparent horizon in $f(T)$ gravity. Bamba et al. (2012a) studied the
finite time singularities, Little Rip, Pseudo-Rip cosmologies and
thermodynamics for the apparent horizon bounded universe for this
gravity.

The nonlinear electrodynamics (NLED) has gained an increasing
revival during last years. This was firstly proposed by Born and
Infeld (1934) who determined an electron of finite radius. After
this achievement, the effects of NLED have been studied in several
papers. De Lorenci et al. (2002) investigated the consequences of
NLED which indicate a universe in the radiation phase. Novello et
al. (2007) determined three different phases of the universe,
bounce, matter and DE phases using NLED. C\^{a}mara et al. (2004)
derived the general nonsingular solution supported by a magnetic
field plus a cosmic fluid and also a non-vanishing vacuum energy
density, which may exhibit the inflationary dynamics of the
universe. Nashed (2011) constructed regular charged spherically
symmetric solutions with NLED coupled to teleparallel theory of
gravity. Bandyopadhyay and Debnath (2011) considered a universe with
magnetic field and matter in NLED and checked the validity of GSLT
for magnetic universe bounded by Hubble, apparent, particle and
event horizons. They concluded that the GSLT violates initially but
holds for later times.

This work provides a motivation to consider a universe with matter,
magnetic field and DE contributions to the energy-momentum tensor.
In this paper, we assume matter in the form of dust and magnetic
field in $f(T)$ gravity, while torsion serves as the DE component.
We evaluate EoS as well as deceleration parameters to explore the
accelerated expansion of the universe. We also check the validity of
the GSLT in this scenario. The format of the paper is as follows: In
section \textbf{2}, we present the preliminaries of the generalized
teleparallel gravity along with NLED for FRW universe. Section
\textbf{3} is devoted to construct some cosmological parameters and
the rate of change of total entropy in the universe for Hubble and
event horizons. We discuss all these results by constructing $f(T)$
models for polelike and power-law forms of scale factor. The
validity of GSLT in this scenario is investigated in section
\textbf{4}. The last section summarizes all the results.

\section{Basics of $f(T)$ Gravity and NLED}

In this section, we provide the basic formulation of the generalized
teleparallel gravity and NLED.

\subsection{Generalized Teleparallel Gravity}

The Riemann-Cartan spacetime is the general structure that possesses
both curvature and torsion tensors. There are two main subclasses of
this spacetime, i.e., the Riemannian spacetime and the
Weitzenb\"{o}ck spacetime. The torsion tensor becomes zero in the
Riemannian spacetime due to symmetric properties of the Levi-Civita
connection defined by the metric tensor. Using this connection with
scalar curvature $R$, Einstein theory of gravity and its
modifications are formed such as
$f(R),~f(R,\mathcal{G}),~f(R,\mathcal{T})$ (De Felice and Tsujikawa
2010; Harko et al. 2011; De Felice et al. 2011) etc. gravity
theories, where $\mathcal{G}$ and $\mathcal{T}$ are the Gauss-Bonnet
invariant and trace of the energy-momentum tensor. By setting
curvature tensor zero, Weitzenb\"{o}ck spacetime is obtained, which
gives rise to TPG and its generalized forms $f(T)$ (Bengochea and
Ferraro 2009; Yang 2011; Ferraro and Fiorini 2011; Myrzakulov 2011;
Tsyba et al. 2011; Linder 2010) and $f(R,T)$ (Myrzakulov 2012;
Chattopadhyay 2012; Sharif et al. 2012), depending upon the tetrad
field and the torsion scalar $T$.

The basic element in the structure of $f(T)$ gravity is the tetrad
field $h_{a}(x^{\mu})$, where the Latin alphabets
$(a,b,...=0,1,2,3)$ denote the tangent space indices and the
spacetime indices are represented by Greek alphabets
$(\mu,\nu,...=0,1,2,3)$. This field forms an orthonormal basis for
the tangent space at each point $x^\mu$ of the manifold and can be
identified by its components $h_\mu^a$ such that $h_a=h^\mu_a
\partial_\mu$. These components satisfy the following properties
\begin{equation}\label{2}
h^{a}_{\mu}h^{\mu}_{b}=\delta^{a}_{b},\quad
h^{a}_{\mu}h^{\nu}_{a}=\delta^{\nu}_{\mu}.
\end{equation}
The relationship between tetrad field and metric tensor $g_{\mu\nu}$
is given by $g_{\mu\nu}=\eta_{ab}h_{\mu}^{a}h_{\nu}^{b}$, where
$\eta_{ab}=diag(1,-1,-1,-1)$ is the Minkowski metric for the tangent
space. With the help of Weitzenb\"{o}ck connection
(${\Gamma^\lambda}_{\mu\nu}=h_{a}^\lambda\partial_\nu h^{a}_\mu$),
the torsion tensor ${T^\rho}_{\mu\nu}$ and the tensor
${S_\rho}^{\mu\nu}$ are defined as follows (Sharif and Jamil 2006,
2007; Sotirious et al. 2011)
\begin{eqnarray}\label{4}
{T^\lambda}_{\mu\nu}&=&{\Gamma^\lambda}_{\nu\mu}-
{\Gamma^\lambda}_{\mu\nu}=h^{\lambda}_{a}
(\partial_{\nu}h^{a}_{\mu}-\partial_{\mu}h^{a}_{\nu}),\\
\label{5}{S_\rho}^{\mu\nu}&=&\frac{1}{2}({K^{\mu\nu}}_{\rho}
+\delta^{\mu}_{\rho}{T^{\theta\nu}}_{\theta}-\delta^{\nu}_{\rho}{T^{\theta\mu}}_{\theta}),
\end{eqnarray}
and ${K^{\mu\nu}}_{\rho}=-\frac{1}{2}({T^{\mu\nu}}_{\rho}
-{T^{\nu\mu}}_{\rho}-{T_\rho}^{\mu\nu})$ is the contorsion tensor.
 These tensors inherit the antisymmetric property and give the
torsion scalar as $T={S_\rho}^{\mu\nu}{T^\rho}_{\mu\nu}$.

The action of $f(T)$ gravity is given by (Bengochea and Ferraro
2009; Yang 2011; Ferraro and Fiorini 2011; Myrzakulov 2011; Tsyba
2011)
\begin{equation}\label{6*}
S=\frac{1}{2\kappa^2}\int d^{4}x[ef(T)+L_m],
\end{equation}
where $e=\sqrt{-g},~\kappa^{2}=8\pi G,~G$ is the gravitational
constant and $L_m$ is the matter Lagrangian density inside the
universe. The corresponding field equations are obtained by varying
this action with respect to tetrad as
\begin{equation}\label{6}
[e^{-1}\partial_{\mu}(eS_{a}~^{\mu\nu})
+h^{\lambda}_{a}T^{\rho}~_{\mu\lambda}S_{\rho}~^{\nu\mu}]f_{T}
+S_{a}~^{\mu\nu}\partial_{\mu}(T)
f_{TT}+\frac{1}{4}h^{\nu}_{a}f=\frac{1}{2}\kappa^{2}h^{\rho}_{a}T^{\nu}_{\rho},
\end{equation}
where $f_T=df/dT,~f_{TT}=d^{2}f/dT^{2}$ and $T^{\nu}_{\rho}$ is the
energy-momentum tensor of perfect fluid. The flat FRW universe is
described by
\begin{equation}\label{7}
ds^{2}=dt^{2}-a^{2}(t)(dx^{2}+dy^{2}+dz^{2}),
\end{equation}
where $a$ is the time dependent scale factor. The corresponding
tetrad components are $h^{a}_{\mu}=diag(1,a,a,a),$ which satisfy
Eq.(\ref{2}). The modified Friedmann equations are
\begin{eqnarray}\label{8}
12H^2f_T+f&=&2\kappa^{2}\rho_{t},\\\label{9}
48H^2\dot{H}f_{TT}-(12H^2+4\dot{H})f_T-f&=&2\kappa^{2}p_{t},
\end{eqnarray}
where $H=\dot{a}/a$ is the Hubble parameter and $\rho_{t},~p_{t}$
are the total energy density and pressure of the universe, dot
represents derivative with respect to time.

\subsection{Nonlinear Electrodynamics}

The standard cosmological model is successful in resolving many
issues but still there are some issues which remain to be solved.
One of the issues is the initial singularity (big bang) which leads
to a troubling state of affairs because at this point, all known
physical theories break down. It has been claimed that very strong
electromagnetic field might help in avoiding the occurrence of
spacetime singularities. Here we give some general properties of the
nonlinear electrodynamics (Born and Infeld 1934; De Lorenci et al.
2002; C\^{a}mara et al. 2004; Novello et al. 2007; Nashed 2011;
Bandyopadhyay and Debnath 2011) in cosmology and discuss its
particular case. The FRW universe model (\ref{7}) requires an
averaging procedure in electrodynamics to maintain its geometry. For
this purpose, the volumetric spatial average of an arbitrary
qunatity $Y$ is defined by
\begin{equation}\label{41}
\overline{Y}=\lim_{V\rightarrow V_{0}}\frac{1}{V}\int
Y\sqrt{-g}d^{3}x,
\end{equation}
where $g$ is the determinant of the metric tensor, $V=\int
\sqrt{-g}d^{3}x$ and $V_{0}$ stands for a sufficiently large time
dependent volume of the whole space. This procedure sets up the mean
values of the electric $E_{i}$ and magnetic $B_{i}$ fields as
follows
\begin{eqnarray}\label{42}
\overline{E_{i}}=0,\quad \overline{B_{i}}=0,\quad
\overline{E_{i}B_{i}}=0,\quad
\overline{E_{i}E_{j}}=-\frac{1}{3}E^{2}g_{ij},\quad
\overline{B_{i}B_{j}}=-\frac{1}{3}B^{2}g_{ij}.
\end{eqnarray}

We consider the extended Maxwell electromagnetic Lagrangian density
up to second order terms in the field invariants $F$ and $F^{\ast}$
as
\begin{equation}\label{43}
\mathcal{L}=-\frac{1}{4}F+\omega_{0}F^{2}+\eta_{0}F^{\ast2},
\end{equation}
with $F=F_{\mu \nu}F^{\mu \nu}=2(B^{2}-E^{2}),~F^{\ast}\equiv
F^{\ast}_{\mu \nu}F^{\mu
\nu}=-4\textbf{E}\cdot\textbf{B},~\omega_{0}$ and $\eta_{0}$ are
arbitrary constants. The Maxwell term (first term) dominates in the
radiation era while the quadratic terms dominates during very early
epoch of the evolving universe. The corresponding energy-momentum
tensor takes the form
\begin{equation}\label{44}
T_{\mu \nu}=-4\mathcal{L}_{F}F_{\mu}~^{\alpha}F_{\alpha
\nu}+(F^{\ast}\mathcal{L}_{F^{\ast}}-\mathcal{L})g_{\mu \nu},
\end{equation}
where $\mathcal{L}_{F}$ and $\mathcal{L}_{F^{\ast}}$ represent the
partial derivatives of the nonlinear Lagrangian with respect to
field invariants. Using the average values given in Eq.(\ref{42}),
the comparison of the energy-momentum tensor (\ref{44}) with that of
perfect fluid, $T_{\mu\nu}=(\rho+p)u_{\mu}u_{\nu}-pg_{\mu\nu}$,
yields the general form of energy density $\rho$ and pressure $p$ as
\begin{eqnarray}\label{47}
\rho&=&-\mathcal{L}-4E^{2}\mathcal{L}_{F},\\\label{48}
p&=&\mathcal{L}+\frac{4}{3}(E^{2}-2B^2)\mathcal{L}_{F}.
\end{eqnarray}

We assume the case of homogenous electric field in plasma which
gives non-vanishing magnetic field whereas the electric field
rapidly decays and becomes zero. The nonlinear term $F^2$ with only
magnetic field helps to avoid the initial singularity by inducing
the universe to bounce (Novello et al. 2007). The vanishing $E^2$
helps to neglect the viscosity terms in the electric conductivity of
the primordial plasma while its presence removes the bounce which
results a universe with a singular state. Inserting the
corresponding values in Eqs.(\ref{47}) and (\ref{48}), the magnetic
energy density and pressure take the form
\begin{eqnarray}\label{49}
\rho_{B}&=&\frac{1}{2}B^{2}(1-8\omega_{0}
B^{2}),\\\label{50}p_{B}&=&\frac{1}{6}B^{2}(1-40\omega_{0} B^{2}).
\end{eqnarray}
When $\omega_{0}=0=\eta_{0}$, Eqs.(\ref{43}) and (\ref{44}) reduce
to the linear Maxwell electromagnetic Lagrangian and energy-momentum
tensor as follows
\begin{equation}\label{45}
\mathcal{L}=-\frac{1}{4}F,\quad
T_{\mu\nu}=F_{\mu}~^{\alpha}F_{\alpha \nu}+\frac{1}{4}Fg_{\mu\nu}.
\end{equation}
For the Lagrangian with the energy-momentum tensor and the same
assumptions as for nonlinear process, we obtain
\begin{equation}\label{46}
\rho=3p=\frac{1}{2}(E^2+B^2),
\end{equation}
which shows that the universe is composed of ordinary radiations
with positive pressure. For the homogenous electric field case, it
corresponds to $k=1$, yielding $p=\frac{1}{3}\rho=\frac{1}{6} B^2$.

\section{Cosmological Parameters and Thermodynamics}

In this section, we construct the EoS and deceleration parameters as
well as the GSLT for Hubble and event horizons. Jamil et al. (2010)
checked the validity of GSLT for a universe composed of DE
interacting with dark matter and radiation fluid. Karami et al.
(2011) studied GSLT for non-flat FRW universe containing the same
fluids for apparent horizon. Here we assume a universe where the
three generic sources fueled its spatial sections including the
pressureless cold dark matter, DE as modified form of torsion scalar
and NLED. The first two sources relate with the late-time evolution
of the universe. The third component is important for avoiding
initial singularity and behaves like standard radiation field at
later times. Thus these contributions develop the budget of energy
density of the universe according to recent observations (i.e.,
$72.8\%$ is DE, $22.7\%$ is dark matter and $4.5\%$ is ordinary
matter) (Komatsu et al. 2011).

The field equations (\ref{8}) and (\ref{9}) can be written as
\begin{eqnarray}\label{10}
\frac{3H^2}{\kappa^2}=\rho_{t},\quad
-\frac{2\dot{H}}{\kappa^2}=\rho_{t}+p_{t},
\end{eqnarray}
where $\rho_{t}=\rho_{m}+\rho_{B}+\rho_{T},~p_{t}=p_m+p_{B}+p_{T}$.
The subscripts $m,~B$ and $T$ denote the matter, magnetic and
torsion contributions to the total energy density and pressure of
the universe with $\rho_T$ and $p_T$ as
\begin{eqnarray}\label{12}
\rho_T&=&\frac{1}{2\kappa^2}(-12H^2f_T-f+6H^2),\\\label{13}
p_T&=&-\frac{1}{2\kappa^2}(48\dot{H}H^2f_{TT}-(12H^2+4\dot{H})f_T-f+6H^2+4\dot{H}).
\end{eqnarray}
For the sake of simplicity, we take dust like matter, i.e.,
$p_{m}=0$. The corresponding energy conservation equations take the
form
\begin{eqnarray}\label{a}
\dot{\rho}_m+3H\rho_m=0,\\\label{b}
\dot{\rho}_B+3H(\rho_B+p_B)=0,\\\label{c}
\dot{\rho}_T+3H(\rho_T+p_T)=0.
\end{eqnarray}
Equation (\ref{a}) gives
\begin{equation}\nonumber
\rho_{m}=\rho_{m0}a^{-3},
\end{equation}
where $\rho_{m0}$ is an arbitrary constant. Inserting Eqs.(\ref{49})
and (\ref{50}) in (\ref{b}), we obtain
\begin{equation}\nonumber
B=\frac{B_0}{a^2},
\end{equation}
where $B_0$ is an arbitrary constant. This shows that the evolution
of energy density of the magnetic field decays with the expansion of
the universe and corresponds to the early phase for small values of
the scale factor (Novello et al. 2004) as well as to radiation phase
in linear case.

Now we investigate the behavior of the universe inheriting magnetic
field and dust matter with $f(T)$ gravity as the DE source. The EoS
parameter is
\begin{eqnarray}\nonumber
\omega_{t}&=&[-\frac{1}{\kappa^2}(48\dot{H}H^2f_{TT}-
(12H^2+4\dot{H})f_T-f+6H^2+4\dot{H})\\\nonumber
&+&\frac{B^2}{6}(1-40\omega_{0}B^{2})]
[\rho_{m0}a^{-3}+\frac{1}{2\kappa^2}(6H^2-f-12H^2f_T)\\\label{16}&+&
\frac{B^2}{2}(1-8\omega_{0}B^{2})]^{-1}.
\end{eqnarray}
The deceleration parameter is the measure of the cosmic acceleration
of the expanding universe and is given by
\begin{equation}\label{17}
q=-1-\frac{\dot{H}}{H^2}.
\end{equation}
The negative value of $q$ corresponds to the accelerated regime, for
positive $q$ decelerated and $q=0$ leads to constant expansion of
the universe. In the present case, it becomes
\begin{equation}\nonumber
q_t=\frac{1}{2}(1+3\omega_{t}),
\end{equation}
hence
\begin{eqnarray}\nonumber
q_{t}&=&\frac{1}{2}[1+3\{-\frac{1}{\kappa^2}(48\dot{H}H^2f_{TT}-
(12H^2+4\dot{H})f_T-f+6H^2\\\nonumber&+&4\dot{H})+\frac{B^2}{6}(1-40\omega_{0}B^{2})\}
\{\rho_{m0}a^{-3}+\frac{1}{2}(6H^2-f-12H^2f_T)\\\label{19}&+&
\frac{B^2}{2}(1-8\omega_{0}B^{2})\}^{-1}].
\end{eqnarray}
Equations (\ref{16}) and (\ref{19}) represent the general form of
the EoS and the deceleration parameters in terms of $f(T)$. We may
check the behavior of these cosmological parameters for some viable
$f(T)$ models.

It has been interesting to study the GSLT in the context of modified
theories of gravity (Akbar and Cai 2006; Sadjadi 2007; Sheykhi and
Wang 2009; Karami and Khaledian 2011, 2012; Karami and Abdolmaleki
2012; Karami et al. 2012). This law states that the sum of entropy
of total matter inside the horizon and entropy of the horizon does
not decrease with time. Using the first law of thermodynamics, the
Clausius relation is obtained as, $-dE=T_{X}dS_X$, where
$S_{X}=\frac{A}{4G}$ is the Bekenstein entropy, $A=4\pi R_{X}^2$ is
the area of horizon with $X$ as an arbitrary horizon and
$T_{X}=\frac{1}{2\pi R_{X}}$ is the Hawking temperature. Miao et al.
(2011) found that the first law of thermodynamics violates in $f(T)$
gravity due to local Lorentz invariance (Li et al. 2011) which
results in addition a entropy production term $S_{P}$. However, its
validation takes place if $f_{TT}$ is very small and entropy horizon
becomes $S_{X}=\frac{Af_T}{4G}$ with vanishing $S_P$ in this case.
We use the general approach (i.e., independent of $f_{TT}$
condition) to study the GSLT in magnetic $f(T)$ scenario along with
Gibbs' equation (Bandyopadhyay and Debnath 2011; Cai and Kim 2005;
Bamba et al. 2012b). The time derivative of the entropy on the
horizon is
\begin{equation}\label{21}
\frac{dS_{X}}{dt}+\frac{dS_{P}}{dt}=\frac{\pi
R_{X}}{G}(2\dot{R}_{X}f_T+R_X\dot{T}f_{TT}).
\end{equation}
If the condition $f_{TT}\ll1$ does not satisfy, then we have to find
out the entropy production term (Bamba et al. 2012b).

The Gibbs' equation is used to find the rate of change of normal
entropy $S_{I}$ of the horizon
\begin{equation}\label{22}
\frac{dS_{I}}{dt}=\frac{1}{T_X}\left(\frac{dE_{I}}{dt}+p_t\frac{dV}{dt}\right),
\end{equation}
where $E_{I}=\rho_{t} V,~V=\frac{4}{3}\pi R_{X}^3$ is the volume of
the horizon. Inserting the values in Eq.(\ref{22}), it follows that
\begin{equation}\label{23}
\frac{dS_{I}}{dt}=\frac{4\pi
R_{X}^{2}}{T_X}(\dot{R}_{X}-HR_{X})(\rho_t+p_t).
\end{equation}
Combining Eqs.(\ref{21}) and (\ref{23}), we obtain the time
derivative of total entropy for the arbitrary horizon as
\begin{eqnarray}\nonumber
\frac{dS_{X}}{dt}+\frac{dS_{P}}{dt}+\frac{dS_{I}}{dt}&=&\frac{\pi
R_{X}}{G}[2\dot{R}_{X}f_T+R_X\dot{T}f_{TT}+8\pi G
R_{X}^{2}\{\rho_{m0}a^{-3}\\\nonumber&+&\frac{1}{\kappa^2}(4\dot{H}Tf_{TT}+2\dot{H}
(f_T-1))+\frac{2B_0^2}{3a^4}(1-\frac{16\omega_{0}B_0^{2}}{a^4})\}\\\label{24}&\times&(\dot{R}_{X}-HR_{X})].
\end{eqnarray}

The validity of the GSLT
($\dot{S}_{X}+\dot{S}_{I}+\dot{S}_{P}\geq0$) on the horizon of
radius $R_{X}$ for viable $f(T)$ models can be investigated. Here we
discuss two forms of cosmological horizons widely used in literature
(Bak and Rey 2000; Li 2004; Sharif and Jawad 2013).

\subsubsection*{Hubble Horizon}

Let us assume that the boundary of the thermal system of the FRW
universe is occupied by the apparent horizon (Bak and Rey 2000) in
equilibrium state. For the flat FRW, it reduces to the Hubble
horizon with radius $R_{H}$ as
\begin{equation}\label{25}
R_{H}=\frac{1}{H},\quad \dot{R}_{H}=-\frac{\dot{H}}{H^2}.
\end{equation}
Inserting these values $(X\rightarrow H)$ in Eq.(\ref{24}), we
obtain
\begin{eqnarray}\nonumber
\frac{dS_{H}}{dt}+\frac{dS_{P}}{dt}+\frac{dS_{I}}{dt}&=&-\frac{\pi}{GH}[\frac{2\dot{H}}{H^2}f_T
+12\dot{H}f_{TT}+\frac{8\pi
G}{H^2}(1+\frac{\dot{H}}{H^2})
\\\nonumber&\times&\{\rho_{m0}a^{-3}+\frac{1}{\kappa^2}(4\dot{H}Tf_{TT}+2\dot{H}
(f_T-1))\\\label{26}&+&\frac{2B_0^2}{3a^4}(1-\frac{16\omega_{0}B_0^{2}}{a^4})\}].
\end{eqnarray}
This is the rate of change of total entropy of all the the fluids
(dust matter, magnetic and torsion contributions) in the universe
for Hubble horizon.

\subsubsection*{Event Horizon}

The radius of event horizon is given by (Li 2004)
\begin{equation}\label{27}
R_{E}=a\int^{\infty}_{t} \frac{dt}{a},\quad \dot{R}_{E}=HR_{E}-1.
\end{equation}
The convergence of this integral leads to the existence of the event
horizon. Basically, it is the distance of light traveling from
present time to infinity. If Big Rip singularity occurs at some
future time denoted by $t_{s}$, then we must replace $\infty$ by
$t_{s}$. Using Eq.(\ref{27}) in (\ref{24}) and replacing $X$ by $E$,
it follows that
\begin{eqnarray}\nonumber
\frac{dS_{E}}{dt}+\frac{dS_{I}}{dt}+\frac{dS_{P}}{dt}&=&\frac{\pi}{G}(a\int^{\infty}_{t}
\frac{dt}{a})[2(\dot{a}\int^{\infty}_{t}\frac{dt}{a}-1)-12H\dot{H}(a\int^{\infty}_{t}
\frac{dt}{a})\\\nonumber&+&8\pi G(a\int^{\infty}_{t}
\frac{dt}{a})^2((\dot{a}\int^{\infty}_{t}
\frac{dt}{a}-1)-H(a\int^{\infty}_{t}
\frac{dt}{a}))\\\nonumber&\times&\{\rho_{m0}a^{-3}+\frac{1}{\kappa^2}(4\dot{H}Tf_{TT}+2\dot{H}
(f_T-1))\\\label{28}&+&\frac{2B_0^2}{3a^4}(1-\frac{16\omega_{0}B_0^{2}}{a^4})\}].
\end{eqnarray}
This represents the rate of change of total entropy in the universe
for event horizon in equilibrium state and its validity depends upon
the viable $f(T)$ model. Sadjadi (2007) investigated the validity of
GSLT for event horizon in $f(R)$ gravity which also depends upon
some viable $f(R)$ model. In the following, we construct some $f(T)$
models to check the behavior of cosmological parameters
$\omega_{t},~q_{t}$ and validity of GSLT for Hubble and event
horizons in a universe composed of dust, magnetic and torsion
contributions.

\section{$f(T)$ Model: An Example}

Since there are mainly two possible ways of working with
cosmological equations of motion, either postulating a theory with
matter content of the universe and then solving corresponding
equation to discuss the cosmological time behavior of the model
under consideration. Or, vice versa, postulating a theory with
desired time behavior of the model deriving information about the
matter content. Novello et al. (2007) investigated the removing of
initial singularity by NLED and resulted a power-law form of scale
factor in the corresponding scenario. Here we adopt the second
method by assuming the following polelike type scale factor (Sadjadi
2006; Nojiri and Odintsov 2006)
\begin{equation}\label{30}
a(t)=a_{0}(t_{s}-t)^{-h},\quad h>0, \quad t_{s}\geq t
\end{equation}
where $a_{0}$ is the present value of the scale factor. This scale
factor indicates the superaccelerated universe with a Big Rip
singularity at $t=t_{s}$. Using above scale factor, Hubble
parameter, torsion scalar and $\dot{H}$ become
\begin{equation}\label{31}
H=\frac{h}{t_s-t},\quad T=-\frac{6h^2}{(t_s-t)^{2}},\quad
\dot{H}=\frac{h}{(t_s-t)^{2}}.
\end{equation}
Inserting these values in the first equation of modified Friedmann
equations (\ref{10}), we obtain the $f(T)$ model as
\begin{eqnarray}\nonumber
f(T)&=&c_1\left(-\frac{T}{6h^2}\right)^{\frac{1}{2}}+\frac{2\kappa^2\rho_{m0}}{a_{0}^3(3h+1)}
\left(-\frac{6h^2}{T}\right)^{\frac{3h}{2}}+
\frac{\kappa^2B_0^2}{a_{0}^4(4h+1)}\left(-\frac{6h^2}{T}\right)^{2h}\\\label{32}&-&
\frac{8\kappa^2B^4_0
\omega_0}{a_{0}^8(8h+1)}\left(-\frac{6h^2}{T}\right)^{4h},
\end{eqnarray}
which shows the contributions from dust matter and magnetic field
with nonlinear terms of torsion scalar. Here $c_1$ is an integration
constant and can be found through a boundary condition. For this
purpose, Eq.(\ref{8}) can be rewritten as follows
\begin{equation}\nonumber
H^2=\frac{8\pi G}{6f_T}\left(\rho_t-\frac{f}{16\pi G}\right).
\end{equation}
This equation implies that the gravitational constant $G$ has to be
replaced by an effective gravitational constant (time dependent),
$G_{eff}$ for nonlinear $f(T)$ model (Capozziello et al. 2011; Wei
et al. 2012). For a linear $f(T),~G_{eff}$ should reduce to the
present day value of $G$ which yields the condition $f_T(T_0)=1$,
where $T_0=-6H_0^2$ and $H_0$ is the present day value of Hubble
parameter. Applying this condition in model (\ref{32}), we obtain
\begin{eqnarray}\nonumber
c_1&=&12hH_0\left[\frac{h\kappa^2\rho_{m0}}{2a_{0}^3(3h+1)H_0^2}
\left(\frac{h}{H_0}\right)^{3h}+
\frac{h\kappa^2B_0^2}{3a_{0}^4(4h+1)H_0^2}\left(\frac{h}{H_0}\right)^{4h}\right.\\\label{x}&-&\left.
\frac{16h\kappa^2B^4_0
\omega_0}{3a_{0}^8(8h+1)H_0^2}\left(\frac{h}{H_0}\right)^{8h}-1\right].
\end{eqnarray}
Also, the model (\ref{32}) satisfies the condition
$\frac{f}{T}\rightarrow 0$ at high redshift $T\rightarrow\infty$ to
be a realistic model representing accelerated expansion of the
universe. This is consistent with the primordial nucleosynthesis and
cosmic microwave background constraints (Wu and Yu 2010; Karami and
Abdolmaleki 2012).

We check the behavior of the expanding universe along with GSLT for
this model by adopting $z=\frac{a_0}{a}-1$. Inserting the above
values in Eqs.(\ref{16}) and (\ref{19}), we obtain
\begin{eqnarray}\label{33}
\omega_t&=&\frac{20\kappa^2B_0^4\omega_0}{9h^2a_0^8}(1+z)^{\frac{8h+2}{h}}-
\frac{\kappa^2B_0^2}{18h^2a_0^4}(1+z)^{\frac{4h+2}{h}}-\frac{2(3h+2)}{3h},\\\label{34}
q_{t}&=&\frac{10\kappa^2B_0^4\omega_0}{3h^2a_0^8}(1+z)^{\frac{8h+2}{h}}-
\frac{\kappa^2B_0^2}{12h^2a_0^4}(1+z)^{\frac{4h+2}{h}}-\frac{5h+4}{2h}.
\end{eqnarray}
\begin{figure} \centering
\epsfig{file=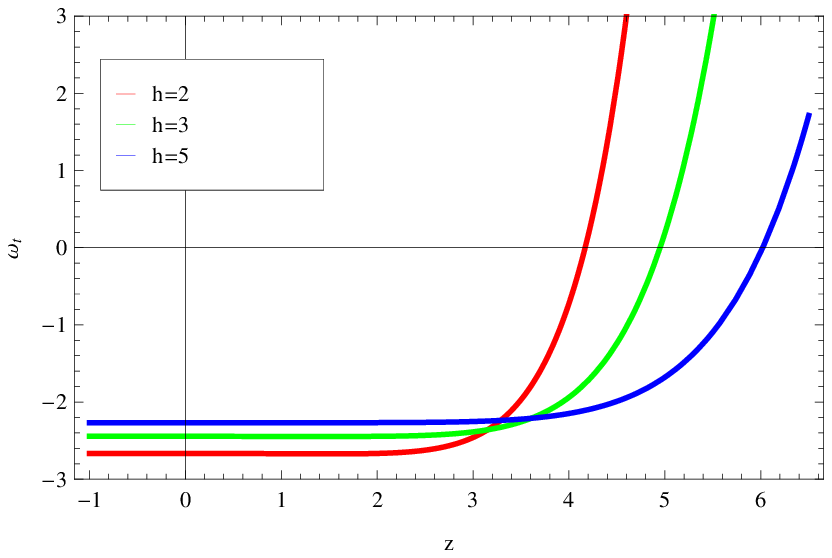,width=.50\linewidth}\epsfig{file=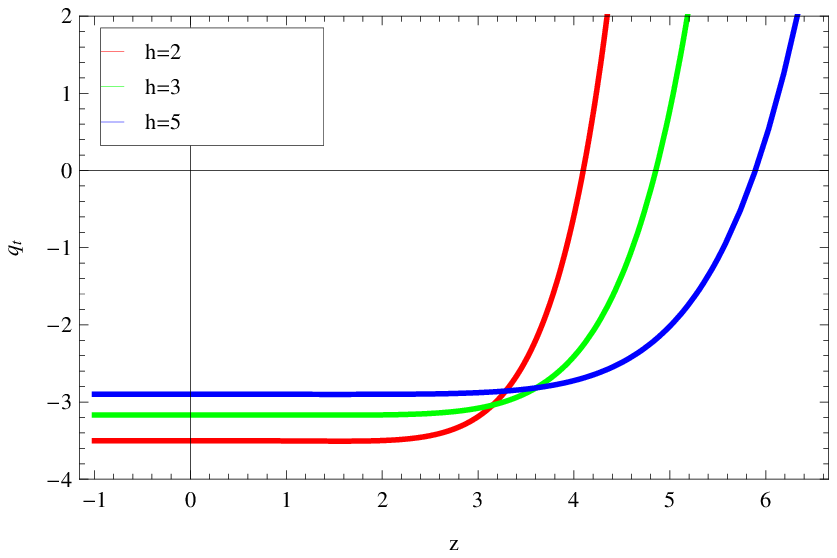,width=.50\linewidth}\caption{Plot
of EoS parameter $\omega_{t}$ (left) and deceleration parameter
$q_{t}$ (right) versus $z$ for polelike type scale factor.}
\end{figure}
The graphical behavior of the cosmological parameters $\omega_{t}$
and $q_{t}$ versus $z$ is shown in Figure \textbf{1}. We use
$a_{0}=1=\kappa^2,~H_0=74.2 Km S^{-1} Mpc^{-1}$ for $h=2,3,5$ and
fix the values $\omega_{0}=0.05,~B_{0}=0.08$ for the magnetic
contribution. In the left graph, $\omega_{t}$ shows the phantom
dominated universe as $z$ decreases for all values of $h$.
Approximately at $z=3.8,4.6,5.6$, the graph shows the crossing of
phantom divide line and converges to phantom era of the expanding
universe which is consistent with the recent observations (Sadjadi
and Vadood 2008). The EoS parameter converges to
$\omega_t=-2.7,-2.5,-2.3$ for $h=2,3,5$ respectively. The range
$z>4.2$ does not correspond to the accelerated phase of the
universe. As we increase the value of $h$, the graph shifts towards
phantom divide line but it crosses the line for higher value of $z$.
The deceleration parameter remains negative for decreasing $z$ as
shown in the right graph. The graph of this parameter becomes
negative in the range, $z<6$ and shows the accelerated expansion of
the universe. For higher values of redshift, roughly $z\geq6$, the
positive behavior of $q_{t}$ indicates the positive decelerated
expansion of the universe. This implies that for decreasing $z$, the
torsion contribution overcomes the magnetic contribution completely.
\begin{figure} \centering
\epsfig{file=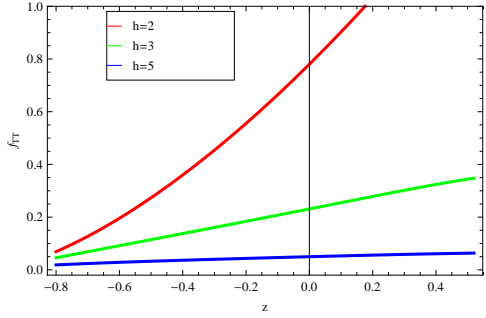,width=.50\linewidth}\caption{Plot of $f_{TT}$
versus $z$ for polelike type scale factor.}
\end{figure}

Now to check the validity of GSLT, we first see the behavior of
second derivative of the model (\ref{32}) given by
\begin{eqnarray}\nonumber
f_{TT}&=&\frac{(1+z)^{\frac{4}{h}}}{36h^4}\left[\frac{3h(3h+2)\kappa^2\rho_{m0}}{2a_{0}^3(3h+1)}(1+z)^3+
\frac{2h(2h+1)\kappa^2B_0^2}{a_{0}^4(4h+1)}(1+z)^4\right.\\\label{32+}&-&\left.
\frac{32h(4h+1)\kappa^2B_{0}^{4}\omega_{0}}{a_{0}^8(8h+1)}(1+z)^8-\frac{c_1}{4(1+z)^{\frac{1}{h}}}\right].
\end{eqnarray}
Its plot versus $z$ is shown in Figure \textbf{2} indicating that
$f_{TT}\ll1$ for $z<0.2$. Thus we take the entropy production term
to zero in Eqs.(\ref{26}) and (\ref{28}). Using Eqs.(\ref{31}) and
(\ref{32}) in (\ref{26}), the rate of change of total entropy in
terms of redshift for the Hubble horizon turns out to be
\begin{eqnarray}\nonumber
\frac{dS_{H}}{dt}+\frac{dS_{I}}{dt}&=&-\frac{\pi
(1+z)^{\frac{3}{h}}}
{Gh^3}\left[\frac{(3h+4)\kappa^2\rho_{m0}}{2a_{0}^3(3h+1)}(1+z)^3+
\frac{4(h+1)\kappa^2B_0^2}{3a_{0}^4(4h+1)}\right.\\\nonumber&\times&\left.(1+z)^4-
\frac{64(2h+1)\kappa^2B_{0}^{4}\omega_{0}}{3a_{0}^8(8h+1)}(1+z)^8-
\frac{c_1}{4h(1+z)^{\frac{1}{h}}}\right]\\\label{35}
&+&\frac{2\pi}{Gh^3}(1+h)(1+z)^{\frac{1}{h}}.
\end{eqnarray}
Figure \textbf{3} (left graph) represents the plot of the rate of
change of total entropy versus redshift keeping the same values of
the constants as in the previous figure. This shows the positive
behavior of $\dot{S}_{H}+\dot{S}_{I}$ for $z>8.2$. As $z$ decreases,
it becomes negative within the range $0\leq z\leq8.2$ and then
converges to zero for $z<0$. Thus the GSLT holds for $z>8.2$ and
$z<0$ in the magnetic $f(T)$ scenario for Hubble horizon.
\begin{figure} \centering
\epsfig{file=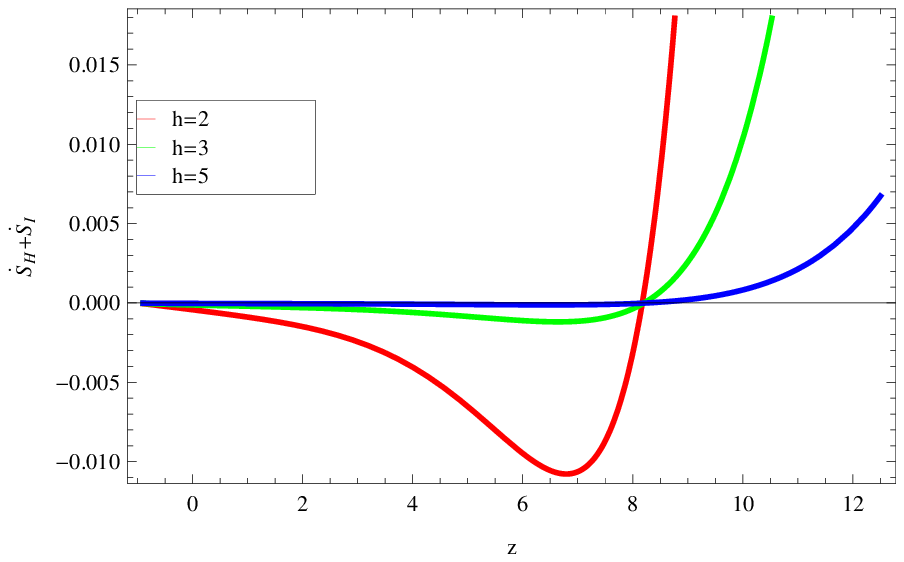,width=.50\linewidth}\epsfig{file=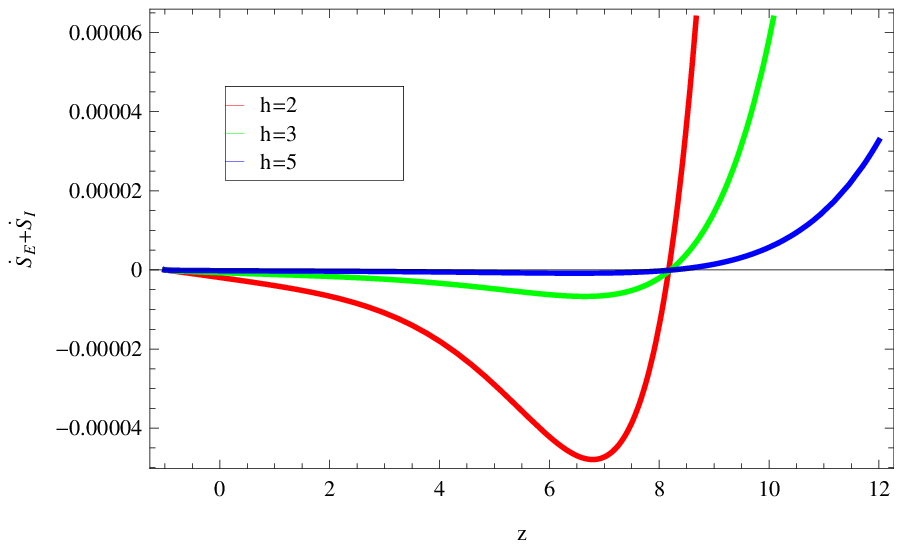,width=.50\linewidth}\caption{Plot
of the rate of change of total entropy versus redshift for polelike
type scale factor. The left graph is for $\dot{S}_{H}+\dot{S}_{I}$
versus $z$ for Hubble horizon and the right graph is for
$\dot{S}_{E}+\dot{S}_{I}$ versus $z$ for event horizon.}
\end{figure}

For the event horizon, inserting Eqs.(\ref{31}) and (\ref{32}) in
(\ref{28}), the rate of total entropy becomes
\begin{eqnarray}\nonumber
\frac{dS_{E}}{dt}+\frac{dS_{I}}{dt}&=&-\frac{\pi
(1+z)^{\frac{3}{h}}}
{Gh(1+h)^2}\left[\frac{(3h+4)\kappa^2\rho_{m0}}{2a_{0}^3(3h+1)}(1+z)^3+
\frac{4(h+1)\kappa^2B_0^2}{3a_{0}^4(4h+1)}\right.\\\nonumber&\times&\left.(1+z)^4-
\frac{64(2h+1)\kappa^2B_{0}^{4}\omega_{0}}{3a_{0}^8(8h+1)}(1+z)^8-\frac{c_1}{4h\kappa^2
(1+z)^{\frac{1}{h}}}\right]\\\label{36}&+&\frac{2\pi
h}{G(1+h)^3}(1+z)^{\frac{1}{h}}.
\end{eqnarray}
The plot of $\dot{S}_{E}+\dot{S}_{I}$ versus $z$ is shown in Figure
\textbf{3} (right graph). This also represents same behavior of the
total entropy for same range of $z$ as for
$\dot{S}_{H}+\dot{S}_{I}$. The only difference lies in the values of
time derivative of total entropies in the corresponding intervals of
$z$. For the magnetic universe only (Bandyopadhyay and Debnath
2011), the GSLT remains valid for Hubble horizon whereas its
validity is investigated up to a certain level along $z$ for event
horizon.

There is another type of scale factor in the exact power-law form
as, $a(t)=a_{0}(t_s-t)^h$ (Sadjadi 2006; Nojiri and Odintsov 2006;
Setare and Darabi 2012) which is simply obtained by replacing $h$
with $-h$ in Eq.(\ref{30}) and gives the inverse power-law
expansion. Debnath et al. (2012) investigated the validity of GSLT
in general relativity using this scale factor (in the limit
$t\rightarrow t_s-t$) along with some other forms of $a(t)$ without
using the first law of thermodynamics. We analyze the behavior of
$\omega_t,~q_t$ and time derivative of total entropy for the Hubble
and event horizons using the same approach as for polelike scale
factor. The corresponding $f(T)$ model is given by
\begin{eqnarray}\nonumber
f(T)&=&c_2\left(-\frac{T}{6h^2}\right)^{\frac{1}{2}}+\frac{2\kappa^2\rho_{m0}}{a_{0}^3(1-3h)}
\left(-\frac{T}{6h^2}\right)^{\frac{3h}{2}}+
\frac{\kappa^2B_0^2}{a_{0}^4(1-4h)}\left(-\frac{T}{6h^2}\right)^{2h}\\\label{32++}&-&
\frac{8\kappa^2B^4_0
\omega_0}{a_{0}^8(1-8h)}\left(-\frac{T}{6h^2}\right)^{4h},
\end{eqnarray}
where $c_2$ is an integration constant which can be found by
applying the same boundary condition as for polelike scale factor,
and is given by
\begin{eqnarray}\nonumber
c_2&=&12hH_0\left[-\frac{h\kappa^2\rho_{m0}}{2a_{0}^3(1-3h)H_0^2}
\left(\frac{H_0}{h}\right)^{3h}-\frac{h\kappa^2B_0^2}{3a_{0}^4(1-4h)H_0^2}
\left(\frac{H_0}{h}\right)^{4h}\right.\\\label{y}&+&\left.
\frac{16h\kappa^2B^4_0
\omega_0}{3a_{0}^8(1-8h)H_0^2}\left(\frac{H_0}{h}\right)^{8h}-1\right].
\end{eqnarray}
This model does not satisfy the condition for a realistic model (Wu
and Yu 2010; Karami and Abdolmaleki 2012) at high redshift. It may
give some relativistic results for $h<\frac{1}{4}$ but this range
does not represent accelerated expansion of the universe ($h>1$).
However, this model satisfies the condition $f(T)\rightarrow0$ as
$T\rightarrow0$ (Rastkar et al. 2012; Chattopadhyay and Pasqua
2013). By comparing this model with Eq.(\ref{32}), the only
difference is the sign of $h$. Thus omitting the expressions for
$\omega_t,~q_t$ and time derivative of total entropy for the Hubble
and event horizons, we discuss graphically these phenomena.
\begin{figure} \centering
\epsfig{file=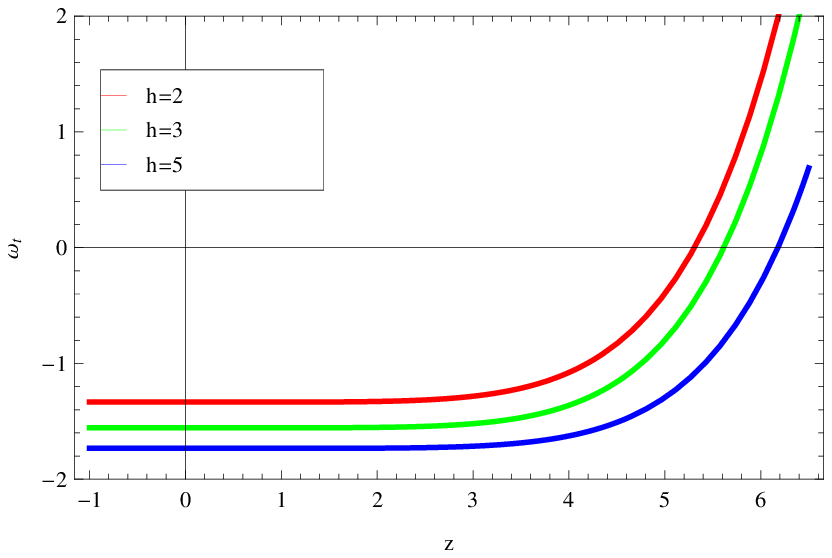,width=.50\linewidth}\epsfig{file=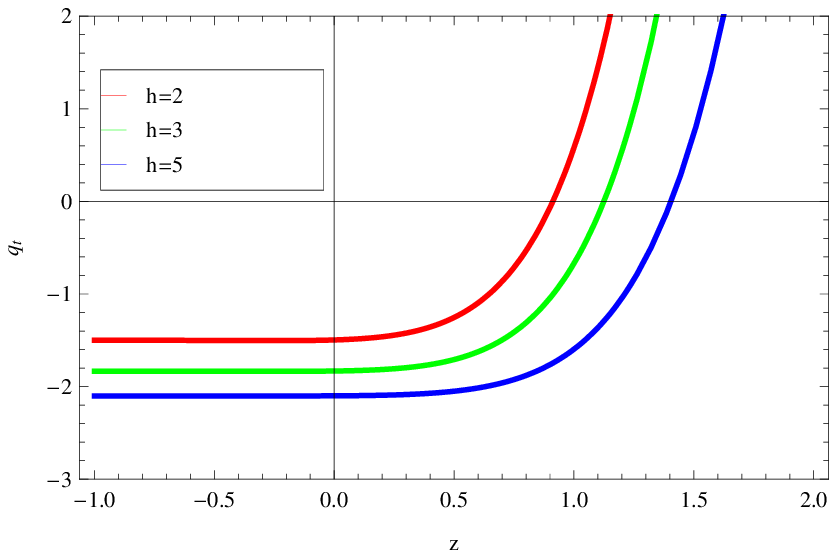,width=.50\linewidth}\caption{Plot
of EoS parameter $\omega_{t}$ (left) and deceleration parameter
$q_{t}$ (right) versus $z$ for power-law scale factor.}
\end{figure}

The EoS and deceleration parameters represent a phantom dominated
accelerated phase of the universe for decreasing values of $z$ as
shown in Figure \textbf{4}. It shows the same behavior of these
parameters as for polelike type scale factor. However, $\omega_t$
crosses the phantom divide line at $z=4.2,4.7,5.4$ and becomes
convergent at $\omega_{t}=-1.3,-1.5,-1.7$ for $h=2,3,5$. As we
decrease the value of $h$, the graph of $\omega_t$ shifts towards
$-1$. For the chosen values of $h$ with $z<7.5$, the deceleration
parameter shows negative behavior. It converges to
$q_t=-1.5,-1.85,-2.1$ for $z<3$ and corresponds to the accelerated
expansion of the universe.

The second derivative of model (\ref{32++}) also satisfies the
condition $f_{TT}\ll1$ as shown in Figure \textbf{5}. Thus we take
$S_P=0$ and check the validity of GSLT for Hubble and event horizons
by taking time derivative of the corresponding total entropy using
Eqs.(\ref{26}) and (\ref{28}) in terms of redshift.
\begin{figure} \centering
\epsfig{file=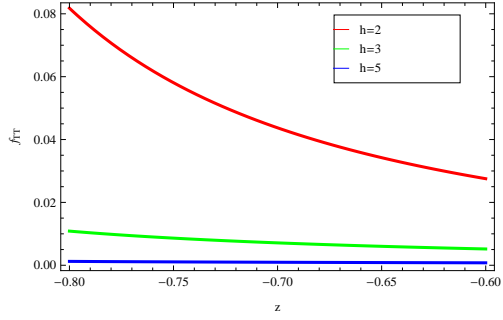,width=.50\linewidth}\caption{Plot of $f_{TT}$
versus $z$ for exact power-law scale factor.}
\end{figure}
The time derivative of total entropy of Hubble horizon shows
positive behavior for $z>-0.05$ and $-0.85$ for $h=2$ and $3$ while
for $h=5$, it remains positive for all values of $z$ as shown in
Figure \textbf{6} (left). Thus GSLT holds in the magnetic $f(T)$
scenario for $h=5$ whereas it violates for $h=2,3$ at $z\leq
-0.05,-0.85$ respectively. In the right graph,
$\dot{S}_{E}+\dot{S}_{I}$ represents positive behavior for all the
values of $z$, showing the validity of GSLT for all chosen values of
$h$ in this scenario.
\begin{figure} \centering
\epsfig{file=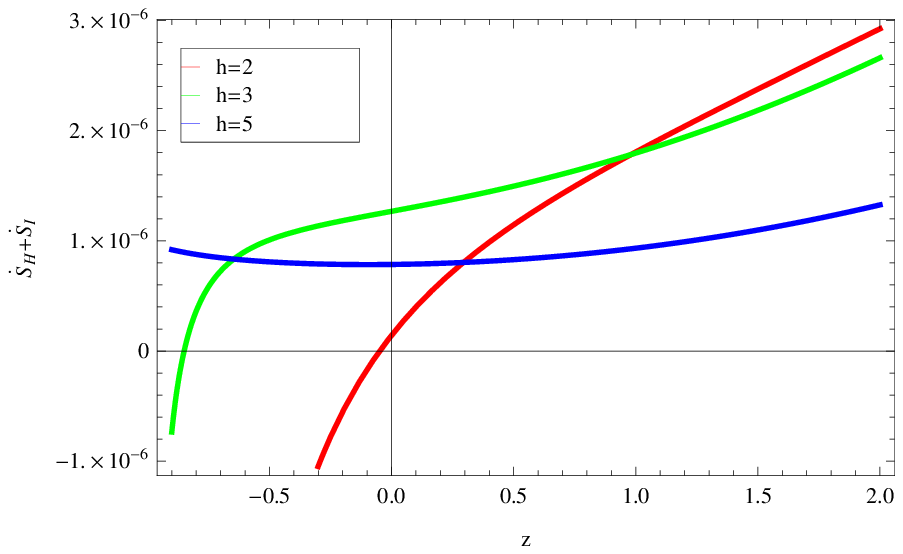,width=.50\linewidth}\epsfig{file=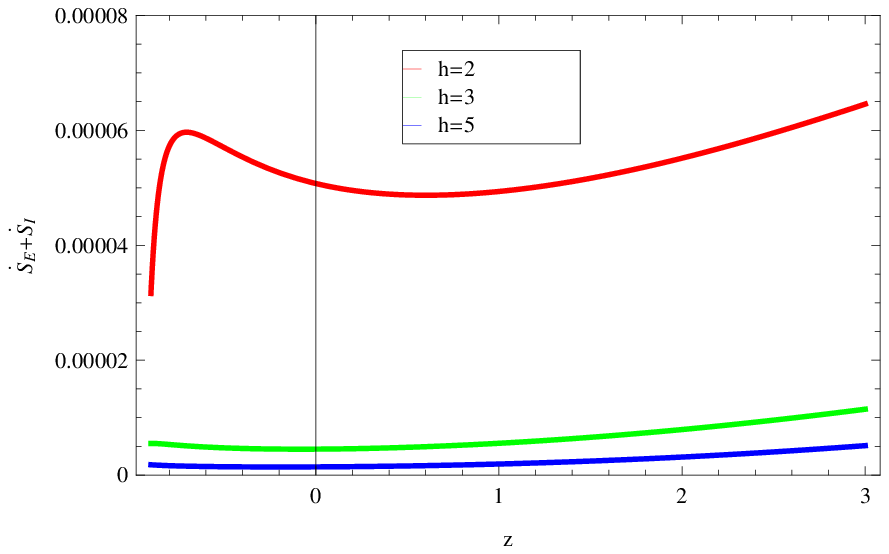,width=.50\linewidth}\caption{Plot
of the rate of change of total entropy versus redshift for exact
power-law scale factor. The left graph is for
$\dot{S}_{H}+\dot{S}_{I}$ versus $z$ for Hubble horizon and the
right graph is for $\dot{S}_{E}+\dot{S}_{I}$ versus $z$ for event
horizon.}
\end{figure}

\section{Concluding Remarks}

We have studied the NLED in the framework of $f(T)$ gravity using
FRW universe containing DE, dust matter and magnetic field
contribution. An averaging procedure is adopted to preserve the
isotropy of spacetime in the NLED. In this scenario, we have
evaluated EoS and deceleration parameters for the total energy
density and pressure of the universe. The time derivative of the
total entropy for the Hubble and event horizons are developed to
investigate the validity of GSLT using horizon entropy and Gibbs'
equation. Dias and Moraes (2005) investigated that torsion affects
the magnetic field only in the topological defect and found that it
spirals up the magnetic field lines along defect axis. The NLED
serves to remove the initial singularity and becomes standard
radiation phase in later times. We have constructed $f(T)$ models
using polelike and power-law forms of scale factor. The graphical
behavior is discussed for some particular model parameters. The
results of the paper are summarized as follows.
\begin{itemize}
\item The cosmological parameters for the first constructed $f(T)$ model by polelike scale factor
represent a phantom dominated universe with acceleration for
$z\leq5.6$ as shown in Figure \textbf{1}. For higher values of $z$,
the expansion rate reduces and magnetic field dominates the torsion
contribution representing a decelerated universe.
\item The time derivative of total entropy for Hubble and event
horizons are plotted versus $z$ to discuss the validity of GSLT for
this model satisfying the condition $f_{TT}\ll1$ (Figure
\textbf{2}). The GSLT holds for $z>8.2$ and $z<0$ for both these
horizons (Figure \textbf{3}).
\item Using the second constructed $f(T)$ model from the exact power-law
scale factor, plots of $\omega_t$ and $q_t$ versus $z$ (Figure
\textbf{4}) indicate the same behavior as the first model.
\item The second model also meets the condition $f_{TT}\ll1$ as
shown in Figure \textbf{5} to discuss the GSLT with the help of
first law of thermodynamics. Figure \textbf{6} shows the positive
behavior of time derivative of the total entropy for Hubble horizon
upto a certain range for $h=2,3$ whereas $h=5$ represents the
validity of GSLT for all values of $z$. For event horizon, the GSLT
is valid for all values of $h$ and $z$.
\end{itemize}

It is interesting to mention here that for the magnetic universe
(Bandyopadhyay and Debnath 2011) only, the time rate of the total
entropy stays positive when $z\geq-0.1$ for event horizon and
becomes negative after this range. On the other hand, in our case,
it remains in the positive region for all values of $z$ in the
magnetic $f(T)$ framework for this horizon with power-law scale
factor. The Hubble horizon shows the similar behavior of the time
derivative of total entropy in both scenarios. For higher values of
redshift, the cosmological parameters indicate a universe where
torsion contribution has become faint as compared to the magnetic
field. It is pointed towards early decelerated phase of the
universe.

\end{document}